\documentclass[12pt]{article}
\usepackage{cite,epsfig,verbatim,rotating,subfigure}
\usepackage{times}
\usepackage{fullpage,setspace}
\usepackage{url}
\usepackage{algorithmic}
\usepackage{algorithm}
\begin{document}

\title{Pyro-Align:Sample-Align based Multiple Alignment system for Pyrosequencing Reads of Large Number}
\author{Fahad Saeed\\
{\small \url{fsaeed2@uic.edu}}\\
\and
{\small\em Department of Biosystems Science and Engineering}
{\small\em ETH Zurich, Switzerland}\\
{\em \&}
\and
{\small\em Department of Electrical and Computer Engineering}\\
{\small\em University of Illinois at Chicago, USA}
}

\maketitle

\singlespacing
\section{Introduction}
\label{sec:intro} Pyro-Align is a multiple alignment program
specifically designed for pyrosequencing reads of huge number.
Multiple sequence alignment is shown to be NP-hard\cite{NP-hard}
and heuristics are desgined for approximate solutions. Multiple
sequence alignment of pyrosequenceing reads is complex mainly
because of 2 factors. One being the huge number of reads, making
the use of traditional heuristics, that scale very poorly for
large number, unsuitable. The second reason is that the alignment
cannot be performed arbitarily, because the position of the reads
with respect to the original genome is important and has to be
taken into account.

Before we indulge in the details of the algorithms itself, a short note on Domain Decomposition and Sample-Align-D algorithm would be useful.

\section {Sample-Align-D and Domain Decomposition}

Sample-Align-D \cite{SaeedKhokhar} algorithm, is a domain
decomposition based technique that is desgined for multiple
alignment of biological sequences on multiprocessor platforms.The
domain decomposition based technique, in addition to yielding
better quality, gives enormous advantage in terms of execution
time and memory requirements. The proposed strategy allows to
decrease the time complexity of any known heuristic of $O(N)^x$
complexity by a factor of $O(1/p)^x$, where $N$ is the problem
size in terms of number and length of sequences, $x$ depends on
the underlying heuristic approach, and $p$ is the number of
processing nodes.We have presented the decomposition strategy as a
parallelcomputing solution. However, the super-linear speed-ups on
multiple processors suggest that the use of the sampling based
decomposition strategy on a single processor systems would also be
able to deliver significant time and space advantages and is thus
used in Pyro-Align. For complete description of the technique the
author refers the reader to \cite{SaeedKhokhar2}.

\section {Pyro-Align ALGORITHM}
I won't give a complete description of the algorithm here but I will try to explain the major components of the system.Hopefully a short summary would help explain the workings of the algorithm in an abstract manner.

The algorithm can be divided into 4 major components:

\begin{enumerate}
\item Overlapping alignment
\item Clustering
\item Pairwise alignments
\item Profile-profile alignments
\end{enumerate}

Each of these components are explained below;

\subsection{Overlapping Alignment}

The first step is to determine where the reads are positioned with respect to the reference genome in question.This step is necessary to ensure that the reads are in the right position with respect to the reference before an actual alignment is executed.Had this step been omitted, there are number of alignments that would be correct but would be inaccurate if analysed in the global context.A read that is not constricted in terms of position, may give the same score (SP score) for the multiple alignment but would be incorrect in context of the reference. To accomplish the task of 'placing' the reads in the correct context with respect to the reference genome we execute overlapping alignment for each read with the reference.

An overlapping alignment ignores the start and end gaps. The overlapping alignment is also known as semi-global alignment because the sequences are globally aligned but the start and end gaps are ignored.Here the start gaps are the gaps that occur before the first 'character' in the sequence and end gaps are that occur after the last character. The overlap alignment can be obtained by modifications to the Needleman-Wunsch algorithm. The alignment thus obtained will potentially provide information of the overlap between the reference genome and the respective read.

The modification to the basic algorithm is now described. Let the two sequences to be aligned be x and y and $A(i,j)$ present the score of the optimal alignment.Since, we would not like to penalise the starting gaps, we modify the dynamic programming matrix by intialising the first row and first coloumn to be zero. The gaps at the end are also not to be penalised. Let $A(i,j)$ represent the optimal score of $x_{1},\cdots,x_{i}$ and $y_{1},\cdots,y_{j}$.Then $A(m,j)$ is the score that represents optimally aligning $x$ with $y_{1,\cdots,j}$.The optimal alignment therefore, is now detected as the maximum value on the last row or coloumn.Therefore the best score is $A(i,j)=max_{k,l} (A(k,n),A(m,l))$ and the alignment can be obtained by tracking the path from $A(i,j)$ to $A(0,0)$.

After each read is semi-globally aligned with the reference genome, we obtain set of leading and trailing gaps, with the first character after the gaps is the starting position for the read with respect to the reference genome. The information for these alignments are stored in hashtables that are further used for processing in the clustering.

\subsection{Clustering}

The method followed by most multiple alignments is that a quick similarity measure is done using k-mer counting or some other heuristic mechanism. Thereafter a distance matrix is computed from the pairwise similarities and a tree is constructed from the distance matrix using UPGMA or neighboring joining. The progressive alignment is thus built, following the branching order of the tree, giving a multiple alignment.The briefly explained procedure for multiple alignment, the two steps of distance matrix and tree construction require $O(N^2)$ time each. For huge number of reads, as in our case this similarity measure is not feasible. Therefore, a method was required which would give us the similarity measure but would be linear in terms of time complexity.

To reduce the complexity of the algorithm, we exploit the information that we already know. We know the fact that the reads are coming from the same reference or nearly same reference. This in turn means that the reads which start from the same or near same 'starting' point with respect to the refernce genome are likely to be similar to each other.Therefore, we already have the clustering information or the 'guide tree' from the first step of the algorithm. Our guide tree, or the order in which sequences will be aligned in the progressive alignment is from the starting position of the reads from the first stage. Ofcourse the decomposition of the reads (the subtree of the profiles that we built) doesnt render the reads in the same order as in traditional progressive alignment, but nevertheless the order is more or less the same when the profiles of these reads are aligned.

\subsection{Pairwise Alignment}

The orginal version of the algorithm, required to align the clusters obtained from the second step, using multiple alignment clustalw system.Our experimentation however, suggested that the pairwise alignment of the reads with same ordering information also give reasonable results.Therefore pairwise local alignment using smithwater is executed on these reads(the ordering is still the same as discussed in section 3.2).After this stage, the reads are aligned in pairs such that we have $N/2$ pair of aligned reads.

\subsection{Profile-Profile Alignments}
\begin{figure}[htb]
\begin{center}
\includegraphics[scale=0.5,angle=+90]{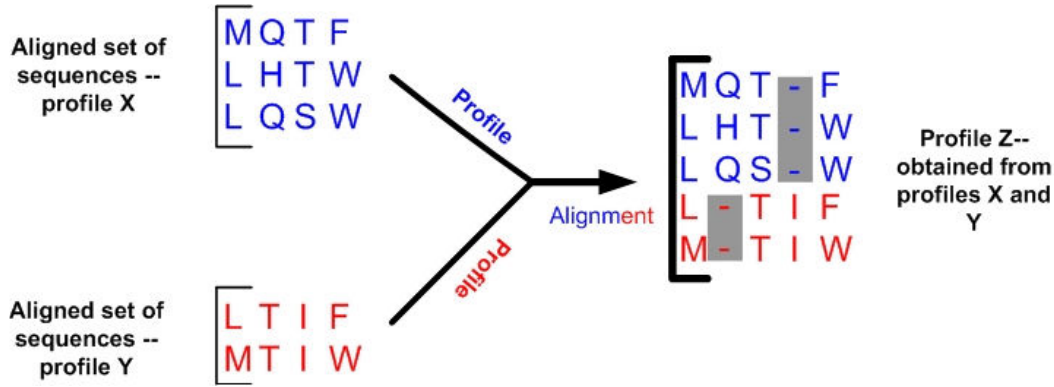}
\caption{\small \label{fig-profile} Two profiles(X and Y) are aligned under the columns constrains, producing profile Z}
\end{center}
\end{figure}

Profile-profile alignments are used to re-align two or more existing
alignments. It is useful for two reasons; one being that the user
may want to add sequences gradually, and second being that the user
may want to keep one high quality profile fixed and keep on adding
sequences aligned to that fixed profile~\cite{clustalw}.
We will take advantage of both of these properties in our domain
decomposition.
In this stage of the algorithm, the $N/2$ pair of aligned reads have to be combined to get a multiple alignment.The profiles of these reads can be aligned sequentially, one  by one, from the clustering information acquired in the first stage. However, we have shown in \cite{SaeedKhokhar2} that the decomposition of the profiles gives a fair amount of time advantages even on single processors. Therefore a hierarchical model similar to \cite{SaeedKhokhar} is implemented in the algorithm.
The model requires that instead of combining the profiles in a sequential manner(one by one), a binary tree is built such that the profiles that are aligned are the leafs of the tree.
Currenltly, the tree is followed for upto 100 clusters(pairs of 2 reads), after which the profiles are  merged in a sequential manner. The reason is to ensure reasonable quality for large number of reads and is extensively discussed in \cite{SaeedKhokhar2}.Profile sum of pairs (PSP) is the function used in Clustalw
\cite{clustalw}, Mafft \cite{mafft} and Muscle \cite{Muscle2}
to maximize Sum of Pairs(SP) score, which in turn maximizes the
Alignment score such that the columns in the profiles are preserved, as depicted in
Fig.~\ref{fig-profile}.

In order to apply pair-wise alignment functions to profiles, a
scoring function must be defined, similar to the substitution
methods defined for pair-wise alignments. One of the most commonly
used profile functions is the sequence-weighted sum of
substitution matrix scores for each pair of amino acid letters.
Let $i$ and $j$ be the amino acid, $p_i$ the background
probability of $i$, $p_{ij}$ the joint probability of $i$ and $j$
aligned to each other, $S_{ij}$ the substitution matrix being
used, $f^x_i$ the observed frequency of $i$ in column $x$ of the
first profile, $x_G$ the observed frequency of gaps in that
column. The same attributes are assumed for the profile $y$. Then
PSP score can be defined as in ~\cite{amino} and \cite{Muscle2}:

\begin{equation}
S_{ij}=\emph{log} (p_{ij}/p_i p_j)
\end{equation}

\begin{equation}
PSP^{xy}= \sum_{i}\sum_{j} f^{x}_{i} f^{y}_{j} \emph{log} (p_{ij}/p_i p_j)
\end{equation}

For our purposes, we will take advantage of PSP functions based on
200 PAM matrix \cite{200PAM} and the 240 PAM VTML matrix
\cite{VTML}. Some multiple alignment methods implement different
scoring functions such as Log expectation (LE) functions, but for
our purposes PSP scoring suffices.Profile functions have evolved to be quite complex and good discussion on these can be found at \cite{Muscle2} and \cite{Muscle3}.We use the profile functions from the clustalw system.

\subsection{Complexity Analysis and Results}
The complexity presented in the section are only approximate. A more rigourous time and space complexity analysis is still required. N is the number of reads, L the averge length of the read and $L_g$ the length of the genome.

The breakdown of time complexity for the algorithm is presented below:
\begin{enumerate}
\item $O(N \times L^2)$ \item $O(N \times L)$ \item $O(N \times
L^2)$ \item $O(N \emph{log}N \times L_g^2)$
\end{enumerate}

The timing for the algorithm is also verified by the experiments
conducted. The time required for aligning 2000 reads on a desktop
computer (Intel dual quad-core Xeon(R) each running at 2.66GHz
with 16GB of RAM and 4096Kb cache size and Linux Redhat with
kernel 2.6.18-92.1.10.e15) using sequential clustalw was observed
to be around 6 hours and 36 minutes. The time observed for
decomsposition based Pyro-Align was around 13 minutes.

\section{Installations \& Running Instructions}

Please refer to the README file in the distibution of Pyro-Align.

\bibliography{mybib}
\bibliographystyle{ieeetr}
\end{document}